\documentclass{iaus}
\usepackage{graphicx}
\newbox\grsign \setbox\grsign=\hbox{$>$}
\newdimen\grdimen \grdimen=\ht\grsign
\newbox\laxbox \newbox\gaxbox
\setbox\gaxbox=\hbox{\raise.5ex\hbox{$>$}\llap
        {\lower.5ex\hbox{$\sim$}}}\ht1=\grdimen\dp1=0pt
\setbox\laxbox=\hbox{\raise.5ex\hbox{$<$}\llap
        {\lower.5ex\hbox{$\sim$}}}\ht2=\grdimen\dp2=0pt
\newcommand{\la}{\mathrel{\copy\laxbox}}
\newcommand{\ga}{\mathrel{\copy\gaxbox}}

\title[Protoclusters and ULXs] 
{Protostellar mergers in protoclusters and the origin 
of ultra-luminous X-ray sources}

\author[Roberto Soria]   
{Roberto Soria$^1$}%

\affiliation{$^1$Harvard-Smithsonian Center for Astrophysics,
60 Garden st, Cambridge, MA 02138, USA 
\break email: rsoria@cfa.harvard.edu}

\pubyear{2005}
\volume{230}  
\pagerange{000-000}
\date{10 Sept 2005}
\jname{Populations of High Energy Sources in Galaxies}
\editors{E.~J.~A.~Meurs \& G.~Fabbiano, eds.}
\begin{document}

\maketitle

\begin{abstract}
I suggest that stellar coalescence in mid-size
protoclusters ($M \sim 10^{3.5}$--$10^{4.5} M_{\odot}$) 
is a possible scenario 
for the formation of ultraluminous X-ray sources (ULXs). 
More massive super-star-clusters are not needed, 
since the most likely ULX mass range  
is only $\sim 30$--$200 M_{\odot}$; 
in fact, they are very rarely found at 
or very near ULX positions. 
Protostellar envelopes and gas accretion 
favour captures and mergers in dense cores 
of embedded clusters. 
Moreover, protoclusters with masses 
$\sim 10^{3.5}$--$10^{4.5} M_{\odot}$ 
are likely to disperse quickly into loose OB associations, 
where most ULXs are found. Sufficiently high 
protostellar density may be achieved when 
clustered star formation is triggered 
by galaxy collisions and mergers. 
Low metallicity may then be necessary to ensure 
that a large fraction of the stellar mass ends up 
in a black hole. In this scenario, most ULXs are naturally 
explained as the extreme end of the high-mass 
X-ray binary population.
\end{abstract}

\firstsection 
\section{Introduction: young or old black holes in ULXs?}
Ultraluminous X-ray sources (ULXs) are accreting compact 
objects with an apparent X-ray luminosity 
$L_{\rm x} > 10^{39}$ erg s$^{-1}$, i.e., higher 
than the Eddington limit of a Galactic stellar-mass 
black hole (BH). Their nature is still hotly debated. 
Three fundamental, unsolved questions are: {\em a)} 
are these sources beamed towards us, or truly ultra-luminous? 
{\em b)} in the latter case, are they emitting above 
their classical Eddington limit, or are they more massive 
than stellar-mass BHs (i.e., $M \ga 30 M_{\odot}$)? 
{\em c)} if the accreting BHs are indeed more massive 
than ``typical'' stellar remnants, were they formed 
in recent star-formation processes, or are they old relics
from the early Universe?

I shall not review here the different arguments in favour 
or against the alternative scenarios in questions {\em a)} 
and {\em b)} (see Miller \& Colbert 2004 for a review).
I shall instead assume for the sake 
of this discussion that most ULXs are not significantly beamed 
sources, and that they do not significantly violate 
the Eddington limit. Hence, I shall accept that ULXs 
in nearby galaxies are powered by accreting BHs 
more massive than those found in our Galaxy  
(``intermediate-mass BHs'', IMBHs, 
with masses $M \ga 30 M_{\odot}$). As for the 
third argument, it has become clear that most ULXs 
(in particular, those brighter than $\approx 3 \times 10^{39}$ 
erg s$^{-1}$) are found in star-forming galaxies, 
not in ellipticals (Irwin et al.~2004; Swartz et al.~2004). 
Colliding or merging galaxies (e.g., the Antennae) 
contain a large number of ULXs, 
often associated with starburst regions, and typical 
stellar populations around ULXs tend to be 
young ($\sim 10$--$50$ Myr). The X-ray luminosity 
function of accreting sources above $10^{39}$ erg s$^{-1}$
is also consistent with ULXs being the bright end 
of the high-mass X-ray binary distribution, 
normalized to the star formation 
rate (Gilfanov et al.~2004). This is not absolute proof that 
the accreting BHs are themselves young: for example, 
they could be old Population-III remnants that 
have recently captured a young donor star while
crossing a dense star-forming region. However, 
the simplest scenario we need to investigate is 
that both the BH progenitors and the donor stars 
are co-eval. 

\section{The super-star-cluster scenario and its shortcomings}

Runaway stellar mergers of main-sequence O stars 
in the collapsed core 
of a young super-star-cluster ($M \sim 10^6 M_{\odot}$, 
size $\sim 1$ pc) have been proposed as a viable  
mechanism to produce a stellar object with a mass up 
to $\sim 1000 M_{\odot}$ at the cluster center, 
which would then collapse into an IMBH 
(Portegies Zwart \& McMillan 2002; G\"{u}rkan et al.~2004).
This process has been used to explain, 
for example, the brightest ULX in the irregular 
starburst galaxy M\,82 (Portegies Zwart et al.~2004).
However, this scenario cannot be used to explain 
most other ULXs. In fact, 
very few of them are inside a massive, compact star cluster.
In a few cases (e.g., the Antennae), 
there are super-star-clusters nearby, 
but the X-ray source is displaced 
by $\sim 100$--$300$ pc. In most other cases, 
there are no super-star-clusters, just OB associations, 
or a group of a few OB stars near the ULX position.

If ULXs were born in compact, massive clusters, 
what happened to them? Was the BH expelled from 
the cluster (unlikely, if it is an IMBH), 
or, more likely, has the parent cluster already dissolved? 
Dispersion of most young star cluster into expanding 
OB associations is seen in the Antennae, on timescales 
$\sim 10^7$ yr (Fall et al.~2005).

I suggest that the runaway merger scenario 
ought to take into account the following 
two basic {\it observational constraints}:
\begin{itemize}
\item the X-ray luminosity distribution has a cut-off 
at $\approx 3 \times 10^{40}$ erg s$^{-1}$ (Gilfanov et al.~2004), 
with only very few sources brighter than that. 
This suggests that the required mass range 
for the accreting IMBHs is only $\sim 30$--$200 M_{\odot}$. 
More massive IMBHs ($M \sim 10^3 M_{\odot}$) 
have been invoked (Miller et al.~2004) 
based on the detection of X-ray spectral features 
interpreted as BH mass indicators. 
In my opinion, those arguments are not 
convincing (see, e.g., Gierli\'{n}ski \& Done 2004  
for an alternative explanation 
of the thermal component at $kT \approx 0.15$ keV) 
and there is no reason to invoke such massive IMBHs.
\item most ULXs are located in OB associations, 
with sizes $\sim 100$ pc and 
$M \sim 10^{3.5}$--$10^{4.5} M_{\odot}$, rather than 
inside compact clusters with sizes $\la$ a few pc 
and $M \sim 10^{5}$--$10^{6} M_{\odot}$.
\end{itemize}

\section{A protocluster scenario}

I propose the following two ingredients to reconcile 
the merger scenario with the two observational 
constraints mentioned in Section 2:

\subsection{Protostellar mergers inside an embedded cluster} 
In the super-star-cluster scenario, 
the timescale available for stellar mergers is $\la 3$ Myr (lifetime 
of a main-sequence O star). If we require the collisions 
to occur already in the protocluster stage, the time available 
is only $\la 0.3$ Myr. Such timescales may still be long enough 
to allow runaway core collapse of mid-size clusters, 
with masses $\la 10^5 M_{\odot}$, 
for stellar velocity dispersions $\la 10$ km s$^{-1}$ 
and central densities $\sim 10^5$--$10^6$ stars pc$^{-3}$ 
(Soria 2005). More significantly, the coalescence rates 
are increased by a few orders of magnitude when 
they involve interactions between protostars, 
surrounded by envelopes or disks, with radii up 
to a few hundred AU, i.e., $\ga 1/10$ of typical 
separations between protostars in a cluster core 
(Bally \& Zinnecker 2005; Elmegreen \& Shadmehri 2003). 
In disk- or envelope-assisted interactions, 
angular momentum of the interacting stars 
can be efficiently dissipated by viscous 
processes and envelope/disk ejection. This favours 
the formation of massive binary protostars. 
Subsequent orbital decay and final coalescence are also 
strongly enhanced during the embedded cluster 
phase: gas accretion onto the protostars leads to orbital 
shrinking and further dissipative interactions with 
the circumstellar material (Bally \& Zinnecker 2005; 
Bonnell \& Bate 2002).

\subsection{Mid-size protoclusters, not super-star-clusters}
Studying the possible formation of a $10^3 M_{\odot}$ IMBH 
in a super-star-cluster may give us 
clues on globular cluster evolution, but 
may not be relevant to the observed ULX population, 
which is generally located in smaller, unbound 
OB associations. However, protoclusters 
in the mass range $\sim 10^{3.5}$--$10^{4.5} M_{\odot}$, 
with central densities $\sim 10^6$ protostars pc$^{-3}$
may be more suitable.
We speculate that these protoclusters can be 
dense enough to allow stellar coalescence 
up to the required IMBH progenitor masses 
(perhaps $\sim 100$--$400 M_{\odot}$). At the same 
time, they are small enough that they tend 
not to survive the embedded phase, evolving 
into unbound OB associations.
There are two main reasons why protoclusters 
in this mass range may be less likely to survive 
into a bound cluster (Kroupa \& Boily 2002; however, 
Fall et al.~2005 argue instead that the survival rate 
within the first $\sim 10^7$ yr is mass-independent):
\begin{itemize}
\item they are massive enough to contain 
many O stars, which ionize all 
the cluster gas; but at the same time, 
they are not massive enough 
to retain the ionized gas ($T_{\rm gas} \sim 10^4$ K 
corresponds to $c_{\rm s} \sim 10$ km s$^{-1}$, 
larger than the escape velocity from the cluster);
hence, they may evaporate ``explosively'' 
(Kroupa \& Boily 2002); 
\item if an IMBH progenitor is produced  
via stellar coalescence in the protocluster 
core, for example by the final merging 
of two $100$-$M_{\odot}$ stars, 
the gravitational energy released by the merger 
($\ga 10^{51}$ erg: Bally \& Zinnecker 2005) 
may be larger than the binding energy 
of the protocluster.
\end{itemize}

Thus, I speculate that if stellar coalescence 
occurs in mid-size protoclusters (rather than 
super-star-clusters), it will be easier 
to explain a population of IMBHs with masses 
$\sim 30$--$200 M_{\odot}$ observed in 
OB associations (leftover of the dispersed 
parent protoclusters). An additional 
advantage of this scenario is that the formation
of ULX progenitors would be essentially 
the same physical process as the formation 
of the progenitors of BH high-mass X-ray binaries 
such as Cyg X-1. This class of massive binary systems 
are also thought to originate from the coalescence 
of less massive protostars inside embedded clusters 
(Bally \& Zinnecker 2005). 
This would be consistent with the observational finding 
that ULXs may simply be the {\it upper end of the high-mass 
X-ray binary distribution}.

\section{Open problems: from massive star to massive BH} 
Whatever the stellar coalescence scenario 
(super-star-clusters or mid-size protoclusters), 
forming a very massive star in the cluster 
core is not enough to have a ULX yet: first, 
the star has to collapse into a sufficiently
massive BH. The final mass of an O star before 
core collapse is generally much less 
than the initial mass, due to 
stellar wind losses.
For example, at solar metallicity, 
a $120 M_{\odot}$ star explodes as a supernova 
with a core mass only $\approx 20 M_{\odot}$ 
(e.g., Vanbeveren 2004).
{\it Low metallicity} reduces this problem, allowing for 
the formation of more massive remnants, for two reasons.
Firstly, mass loss in the stellar wind is much reduced: 
$\dot{M} \sim Z^{0.86}$ for $10^{-2} \la Z/Z_{\odot} \la 1$ 
(Vink \& de Koter 2005). Secondly, at sub-solar (but not primordial) 
metallicity, all stars with masses $\ga 40 M_{\odot}$ 
are thought to collapse directly into a BH (Heger et al.~2003). 
Weaker winds and direct BH collapse of the progenitor 
may be the reason why ULXs appear to prefer 
low-metallicity environments, 
as originally suggested by Pakull \& Mirioni (2002). 

An additional requirement for ULX formation 
is that the BH has a Roche-lobe-filling 
companion star able to transfer  
$\ga 10^{-6} M_{\odot}$ yr$^{-1}$. 
This is a comparatively minor problem: 
such steady mass transfer rates are possible 
for donor stars $\ga 10 M_{\odot}$, over 
their nuclear timescale (a few $10^6$ yr)   
(Rappaport et al.~2005).
Direct BH collapse of the primary may increase 
the likelihood of retaining a companion star, 
which will later become the donor.

As an aside, I suggest that future optical measurements 
of their {\it proper motion distribution} 
may reveal whether ULXs are powered by beamed 
or super-Eddington stellar-mass BHs (in which case 
they would tend to have higher velocities, 
from SN kicks), or are more massive 
systems, formed via direct BH collapse (in which 
case no kick is expected).

\section{Conclusions}
I argue that one can explain the vast majority of ULXs  
with accreting BHs with masses $\sim 30$--$200 M_{\odot}$. 
Higher masses are probably not required. Thus, we do not need 
to invoke runaway merger processes in super-star-clusters 
($M \sim 10^6 M_{\odot}$), 
which are not generally found at or near ULX locations.
Coalescence of a few massive stars in a mid-size cluster 
($\sim 10^{3.5}$--$10^{4.5} M_{\odot}$) may be enough 
to explain the BH progenitors. 

Moreover, I suggest 
that the merger process should occur in the protocluster 
stage: it is much easier to capture and merge protostars 
(surrounded by large disks or envelopes) 
than main-sequence stars, for the same stellar density 
and velocity dispersion. Core densities $\sim 10^6$ (proto)stars 
pc$^{-3}$ are required for mergers to become significant. 
Such high densities are probably achieved when 
clustered star formation is triggered by molecular cloud 
collisions in merging galaxies (e.g., Keto et al.~2005).
This may explain the preferential association of ULXs 
with tidally-disturbed environments.

Protoclusters in the $10^{3.5}$--$10^{4.5} M_{\odot}$ 
mass range are known to disperse quickly, 
evolving into OB associations 
rather than bound clusters. This is 
in agreement with the fact that most ULXs 
are found in mid-size stellar groups or OB associations 
rather than massive, bound clusters.
Low metal abundance may be the other additional 
ingredient, ensuring low mass-loss in the stellar 
wind of the progenitor, followed by its direct BH collapse. 
In summary, perhaps the most important keys to understand 
ULX formation will come  
from infrared, sub-mm and radio studies 
of massive star formation in embedded clusters.


\end{document}